\documentstyle[aps,preprint]{revtex}
\draft

\author{ Zafar Ahmed \\
Nuclear Physics Division, Bhabha Atomic Research Centre \\
Trombay, Bombay 400 085, India \\ zahmed@apsara.barc.ernet.in }
\title
{P, T, PT, and CPT invariance of Hermitian Hamiltonians}
\date{\today}
\begin{document}
\maketitle
\begin{abstract}
Currently, it has been claimed that certain Hermitian Hamiltonians have
parity (P) and they are PT-invariant. We propose generalized definitions
of time-reversal operator (T) and orthonormality such that all Hermitian
Hamiltonians are P, T, PT, and CPT invariant. The PT-norm and CPT-norm are
indefinite and definite respectively. The energy-eigenstates are either E-type
(e.g., even) or O-type (e.g., odd). C mimics the charge-conjugation symmetry
which is recently found to exist for a non-Hermitian Hamiltonian. For a
Hermitian Hamiltonian it coincides with P.
\end{abstract}
\vspace {.2 in}
\par The Hamiltonians which are invariant under the joint transformation of
Parity $(x\rightarrow -x)$ and Time-reversal $(i \rightarrow -i)$ are called
PT-invariant. It has been conjectured [1] that such Hamiltonians possess real
discrete energy-eigenvalues provided the PT symmetry is unbroken.
PT-symmetry is called unbroken or exact if the energy-eigenstates are also
simultaneous eigenstates of PT. On the other hand when PT-symmetry is spontaneously
broken the energy-eigenvalues are complex conjugate pairs.
Multipronged investigations supporting this conjecture have been extensively
carried out [1-3]. Consequently, the condition of Hermiticity for a Hamiltonian
to possess real eigenvalues gets relaxed. It is remarkable that it is discrete
symmetries of an Hamiltonian which seem to decide if the eigenvalues will be
real.
\par The real eigenvalues of the PT-symmetric potential can now be found to
be connected to the concept of $\eta-$ distorted inner-product
$\langle \psi| \eta \psi \rangle$ [4] which culminated in 50s-60s into
the concept of pseudo-Hermiticity [5] of a Hamiltonian
\begin{equation}
\eta H \eta^{-1} =H^\dagger,
\end{equation}
wherein it is known [5] that the distinct real eigenvalues have $\eta-$orthogonal
eigenvectors. The complex eigenvalues are known to have zero $\eta-$norm.
Recently, several PT-symmetric potentials have been pointed out to be
P-pseudo-Hermitian [6]. Several other classes of non-Hermitian Hamiltonians
which are both PT-symmetric and non-PT-symmetric Hamiltonians have been argued
to be pseudo-Hermitian under $\eta=e^{-\theta p}$ and $\eta=e^{-\phi(x)}$ [7].
Pseudo-Hermiticity has been found to be more general when non-Hermitian Hamiltonians
have real eigenvalues. However, the notion of PT-symmetry is physically more
appealing which could provide contact with physical systems and situations.
Recently, by constructing $2 \times 2$ pseudo-Hermitian matrices
a new pseudo-unitary group and new ensembles of Gaussian-random matrices have
been proposed [8]. New energy-level-spacing-distribution functions hence obtained [8] are
expected to represent the spectral fluctuations of PT-symmetric systems.
\par A real central potential in three dimensions and real a symmetric (under space reflection)
potentials in one dimension are automatically PT symmetric. It is the real, non-symmetric
potential in one dimension which by admitting real eigenvalues would disallow
a conjecure that Hermitian potentials are PT-symmetric. However,
currently for Hamiltonians of the type ${\cal H}=p^2/(2m)+V(x)$
two fundamental claims have been made : ${\bf C_1
:}$``Hermitian Hamiltonians, ${\cal H}$, have Parity '' and ${\bf C_2 :}$
``Hermitian Hamiltonians, ${\cal H}$, are PT-invariant (converse not true).''[9]
These claims are fundamentally important as they connect the discrete symmetry
( PT-symmetry) of the Hamiltonian to the reality of eigenvalues in the conventional quantum mechanics.
In this Letter, we shall examine, extend and consolidate these claims further.
Let us remark that this PT-symmetry of a Hermitian Hamiltonian
[9] in contrast to the conjecture of Bender and Boettcher [1] essentially
includes the individual P and T invariance of a Hamiltonian.
\par According to the claim ${\bf C_1}$ Hermitian Hamiltonians, ${\cal H}$, have parity
[9]. If ${\cal H}$ is a Hermitian Hamiltonian with an eigenvalue equation
\begin{equation}
{\cal H}|\Psi_n \rangle = E_n |\Psi_n \rangle.
\end{equation}
The completeness and orthonormality of the eigenstates read as
\begin{equation}
\sum_{n=0}^{\infty}|\Psi_n \rangle \langle \Psi_n|= 1,
\langle \Psi_m | \Psi_n \rangle=\delta_{m,n}.
\end{equation}
The parity operator P has been proposed [9] as
\begin{equation}
P=\sum_{n=0}^{\infty} (-1)^n |\Psi_n \rangle \langle \Psi_n|.
\end{equation}
It can be shown that P is involutary and it commutes with ${\cal H}$ and its eigenvalues
are $\pm 1 $ i.e.,
\begin{equation}
P^2=1, [P,{\cal H}]=0, P|\Psi_n \rangle =(-)^n |\Psi_n \rangle.
\end{equation}
For the Hamiltonians of the type $H_S=p^2/2m+V_S(x)$ where $V_S(x)$ is symmetric
under space-reflection this
proposal works well as the eigenstates will be symmetric
(anti-symmetric) as $n$ is even (odd).
Imagine, if $V_S(x)$ is slightly distorted to make it non-symmetric
($V_{N-S}(x)$), this potential will have the same number of bound-states but
now they can no more be classified as even/odd functions of space variable
despite their quantum numbers being even/odd. The claim ${\bf C_1}$ that all
Hermitian Hamiltonian have parity would break down.
\par When there exists a symmetry in the system one can classify the state
and a certain kind of order can be observed. When this symmetry is broken,
we loose the order and would find it difficult to classify the states again.
Thus, we propose a new classification scheme of the states
for the potential $V_{N-S}(x)$ to revive the claim ${\bf C_1}$. We term
the states as Extraordinary-type (E-type)
when the wavefunctions satisfy a condition that $\Psi_n(x=-L) \Psi_n (x=R)>0$
and Ordinary-type (O-type)
when we have $\Psi_n(x=-L) \Psi_n (x=R)<0$ . Here, $L,R$ are the large asymptotic distances on either side of the
potential. Therefore, one can now state that all Hermitian Hamiltonians
have a {\it generalized} parity (4) wherein the states are either E-type or O-type.
The E-type (O-type) of states have even (odd) number of nodes. It may be well
to recall that in Bohr-Sommerfeld or WKB quantization of the Hamiltonian,
the quantum number $n$ is set even and odd alternatively to get the complete
spectrum irrespective of the symmetry of the potential. These methods do seem
to have a {\it generalized} sense built in them. Let us remark that the Hamiltonians of the
type $[p-\phi]^2/(2m)+V(x)$ could be treated as $H_{N-S}$.
\par The most interesting aspect of the $\eta-$norm $(\langle \Psi| \eta \Psi
\rangle)$ [4,5] or PT-norm is its indefiniteness (positive-negative) [3]
($\langle \Psi| P \Psi \rangle)$ as against the positive definiteness of the
usual (unitary, Hermitian) norm $(\langle \Psi|\Psi \rangle).$
Since the norm represents the quantum mechanical probability, an indefinite PT
or $\eta-$ norm is taken to be very seriously. In this regard, a current
proposal [10] that the negativity of the PT-norm indicates a hidden symmetry
which would mimic [11] charge-conjugation symmetry (C) such that CPT-norm is
positive definite is very appealing.
Consequent to this a pseudo-Hermitian (1) $2 \times 2$ matrix Hamiltonian has
been demonstrated to be C,PT,CPT invariant by constructing $P=\eta$, $T=K_0$ and
C in an interesting way [10].
\par It becomes natural to put the claims ($\bf C_{1,2}$) in this more general
perspective for the sake of consistency. Since the potentials considered in [9]
are real and therefore the PT-invariance of Hermitian Hamiltonians is automatic.
We find that the definition of T as $K_0$ [10], if extended to the Hermitian matrix
Hamiltonians, would actually disprove the claim ${\bf C_2}.$
To be both consistent and rigorous, one actually requires a {\it generalized}
definition {\it a la} (4) of an anti-linear, involutary operator associated with time-reversal
symmetry T.
\par To this end, we would like to switch over to matrix notation of eigenstates
$\Psi_n$. Let us recall that we can have three operations over $\Psi_n$ i.e.,
complex-conjugation $(\Psi_n^\ast)$, transpose operation $(\Psi_n ^\prime)$
and both together as $\Psi_n^\dagger$ which denotes the Dirac's bra-vector :
$\langle \Psi_n|=|\Psi_n\rangle^\dagger$. Without loss of generality, we assume
the Hamiltonian to be a $2 \times 2$ matrix with eigenvalue equation as
$H\Psi_n=E_n\Psi_n$ $(n=0,1)$, so we have
\begin{equation}
\Psi_0 \Psi_0^\dagger +\Psi_1 \Psi_1^\dagger =1, \Psi_n^\dagger \Psi_m=\delta_{m,n}.
\end{equation}
The parity operator (4) becomes
\begin{equation}
P=\Psi_0 \Psi_0^\dagger - \Psi_1 \Psi_1^\dagger
\end{equation}
yielding
\begin{equation}
P^2=\left (\Psi_0 (\Psi_0^\dagger \Psi_0) \Psi_0^\dagger -\Psi_1 (\Psi_1^\dagger
\Psi_0) \Psi_0^\dagger-\Psi_0 (\Psi_0^\dagger \Psi_1) \Psi_1^\dagger
+\Psi_1 (\Psi_1^\dagger \Psi_1) \Psi_1^\dagger \right ) = \left ( \Psi_0 \Psi_0^\dagger
+\Psi_1 \Psi_1^\dagger \right ) = 1,
\end{equation}
\begin{equation}
\mbox {and}~~[P,H]=P H-H P=0, P\Psi_n=(-)^n\Psi_n.
\end{equation}
We propose the anti-linear time-reversal operator T as
\begin{equation}
T=UK_0=\left( \Psi_0 \Psi_0^\prime + \Psi_1 \Psi_1^\prime \right) K_0.
\end{equation}
Here, $K_0$ is the complex-conjugation operator i.e., $K_0 (A B+ C D)=
A^\ast B^\ast+ C^\ast D^\ast$.
the operator T is involutary
\begin{equation}
T^2=UK_0UK_0=\left (\Psi_0 \Psi_0^\dagger+\Psi_1 \Psi_1^\dagger \right)=1.
\end{equation}
T commutes with H
\begin{equation}
[T,H]=T H-H T=0, T\Psi_n=\Psi_n.
\end{equation}
Using (7) and (10), we construct PT or TP operators as
\begin{equation}
PT=\left(\Psi_0 \Psi_0^\prime -\Psi_1 \Psi_1^\prime \right) K_0= TP,
\end{equation}
which has the following properties :
\begin{equation}
(PT)^2=1, [PT,H]=0, PT\Psi_n=(-1)^n\Psi_n.
\end{equation}
We now define a general $\chi$-orthonormality and $\chi-$norm as
\begin{equation}
(\chi \Psi_m)^\dagger \Psi_n = C_{m,n}\delta_{m,n}, N_{\chi}=(\chi \Psi)^\dagger \Psi,
\end{equation}
$C_{m,n}$ is indefinite (positive-negative). $\chi$ in the above equation denotes
discrete symmetry operators ( P,T, and PT ) of the Hermitian Hamiltonian,
i.e $[\chi, H]=0.$
Thus for a Hermitian Hamiltonian H we get
\begin{equation}
N_{PT,n}=(-1)^n, n=0,1 ,
\end{equation}
which is indefinite.
The indefiniteness of PT-norm when the Hamiltonian is non-Hermitian, PT-symmetric
(pseudo-Hermitian) has motivated a novel identification of charge-Conjugation
symmetry, C, in order to make the CPT-norm definite [10].
\par In our case when the Hamiltonian is Hermitian, choosing one from $\Psi^\dagger, \Psi^\prime$ and other from $\Psi^\ast,
\Psi$, one can construct only two distinct and nontrivial involutary operators P and T.
One can therefore not associate with a Hermitian Hamiltonian third distinct linear
involutary operator which could possibly be charge-conjugation-operator $C$ such that $C^2=1.$
Notice that by setting $C=P$, we find that $N_{CPT}$ ($\chi=CPT=P^2T=T$)
in (15) is positive definite and Hamiltonian is CPT-invariant i.e.,
[H,CPT]=[H,$P^2$T]=[H,T]=0. \\
Let us re-emphasize that the definition of T assumed as $K_0$
in Ref. [10] fails to prove the T and PT-invariance of a Hermitian matrix
Hamiltonian.
Here, we are able to define T and norm as in Eq.(10) and Eq. (15) respectively which
salvages this problem and one can prove the claimed [9] PT-symmetry of Hermitian
Hamiltonian in general. In the illustration below this point is being brought out. \\
{\bf Illustration : } \\
Let the Hermitian Hamiltonian be modelled as
\begin{eqnarray}
H=\left [ \begin{array}{cc} a & b+ic \\ b-ic & a \end{array}\right],
\end{eqnarray}
The eigenvalues are $E_{0,1}=a \pm \sqrt{b^2+c^2}$ and the normalized eigenvectors are
\begin{eqnarray}
\Psi_0={1\over \sqrt{2}}\left [ \begin{array}{c} e^{i\theta} \\ 1  \end{array}\right],
\Psi_1={1\over \sqrt{2}}\left [ \begin{array}{c} e^{i\theta} \\ -1  \end{array}\right]
\end{eqnarray}
where $\theta=\mbox{tan}^{-1}(c/b).$ Using (7) and (10) P and T can be constructed as
\begin{eqnarray}
P=\left [ \begin{array}{cc} 0 & e^{i\theta} \\ e^{-i\theta} & 0 \end{array}\right],~~~
T=\left [ \begin{array}{cc} e^{2i\theta} & 0\\ 0 & 1 \end{array}\right]K_0,~~~
PT=\left [ \begin{array}{cc} 0 & e^{i\theta}\\  e^{i\theta} & 0 \end{array}\right]K_0
\end{eqnarray}
One can readily verify the involutions : $P^2=1=T^2=(PT)^2$ and commutations
revealing the discrete symmetries of H : [P,H]=[T,H]=[PT,H]=0. One can confirm that
$\Psi_{0,1}$ are also the eigenstates of P,T, and PT. One can see that $N_{PT,n}=(-1)^n$.
However, following the Ref. [10], if we assume T=$K_0$, it can be quickly be seen
that $[T,H] \ne 0 \ne [PT,H]$. The PT-orthonormality as defined in [10] which in matrix
notation reads as $(P K_0 \Psi_0)^\prime \Psi_1$ does not vanish and becomes
complex ! This justifies our definitions of T and $\chi$-orthonormality given in
(10) and (15) respectively.
Let us point out that generalization to $N \times N$ matrix Hamiltonians
is straightforward i.e.,
\begin{equation}
P=\sum_{n=0}^{N} (-)^n\Psi_n \Psi_n^\dagger,
T=\left (\sum_{n=0}^{N} \Psi_n \Psi_n^\prime \right)K_0
\end{equation}
\par Thus, by employing our proposed definitions of T and $\chi$-orthonormality in
(10) and (15) we could establish and demonstrate that Hermitian Hamiltonians are P-symmetric,
T-symmetric, PT-symmetric, CPT-symmetric and the eigenstates are either
E-type or O-type. PT-norm (CPT-norm) is indefinite (definite).
In the light of this work one now requires the definitions of linear (C,P)
and anti-linear operator T when the Hamiltonian is pseudo-Hermitian matrix [8]
possessing  real eigenvalues. As the basis for a pseudo-Hermitian Hamiltonian
is known to be bi-orthonormal $(\Psi,\Phi)$ [5], this gives a possible handle for
constructing one more involutary operator C other than P and T.
Thus found definitions of P, T, and C and orthonormality are expected to be consistent
with the definitions discussed here (7,10,15). In fact, the new definitions
ought to contain the present ones as a special case.
These constructions, however, turn out to be quite elusive presently.
It is instructive to note that matrix notations are not only handy but also
are more transparent and general than Dirac's notations of bras and kets.
This feature brings the present work closer to the discrete symmetries
${\cal P},{\cal T}$ and ${\cal C}$ which are discussed in relativistic field
theory [11].
\section*{References }
\begin{enumerate}
\item C.M. Bender and S. Boettcher, Phys. Rev. Lett. 80 (1998) 5243;
\item M. Znojil, Phys. Lett. A {\bf 259} (1999) 220; {\bf 264} (1999) 108.\\
 C. M. Bender, G. V. Dunne, P. N. Meisinger, Phys. Lett. A {\bf 252} (1999) 253. \\
 H.F. Jones, Phys. Lett. A {\bf 262} (1999) 242. \\
 B. Bagchi and R. Roychoudhury, J. Phys. A : Math. Gen. A {\bf 33} (2000) L1.\\
 G. Levai and M. Znojil, J. Phys. A : Math. Gen. {\bf 33} (2000) 7165.\\
 A. Khare and B.P. Mandal, Phys. Lett. A {\bf 272} (2000) 53.\\
 B. Bagchi and C. Quesne, Phys. Lett. A {\bf 273} (2000) 285. \\
 R. Kretschmer and L. Symanowaski, `The interpretation of quantum mechanical
 models with non-Hermitian Hamiltonians and real spectra', arXive
 quant-ph/0105054 \\
 Z. Ahmed, Phys. Lett. A : {\bf 282} (2001) 343;{\bf 287} (2001) 295;
 {\bf 286} (2001) 231.\\
 R.S. Kaushal, J. Phys. A: Math. Gen, {\bf 34} (2001) L709.\\
 R.S. Kaushal and Parthasarthi, J. Phys. A : Math. Gen. {\bf 35} (2002) 8743.\\
 Z. Ahmed,`Discrete symmetries, Pseudo-Hermiticity and pseudo-unitarity', in
DAE (India) symposium on Nucl. Phys. Invited Talks eds., A.K.Jain and A. Navin,
vol {\bf 45 A} (2002) 172.
\item Z. Ahmed, `A generalization for the eigenstates of complex PT-invariant
 potentials with real discrete eigenvalues' (unpublished) (2001). \\
 M. Znojil, `Conservation of pseudo-norm in PT-symmetric quantum mechanics',
 arXive quant-ph/0104012.\\
 B. Bagchi, C. Quesne and M. Znojil, Mod. Phys. Lett. A{\bf 16} (2001) 2047.\\
 G.S. Japaridze, J. Phys. A : Math. Gen. {\bf 35} (2002) 1709.
\item P.A.M. Dirac, Proc. Roy. Soc. London {A 180} (1942) 1.\\
 W. Pauli, Rev. Mod. Phys. {\bf 15} (1943) 175.\\
 T.D. Lee, Phys. Rev. {\bf 95} (1954)  1329.\\
 S.N. Gupta, Phys. Rev. {\bf 77} (1950) 294.\\ K. Bleuler, Helv. Phys. Act.
 {\bf 23} (1950) 567.
\item
 R. Nevanlinna, Ann. Ac. Sci. Fenn. {\bf 1} (1952) 108; {\bf 163} (1954) 222.\\
 L.K. Pandit, Nouvo Cimento (supplimento) 11 (1959) 157.\\
 E.C.G. Sudarshan, Phys. Rev. {\bf 123} (1961) 2183.\\
 M.C. Pease III, {\it Methods of matrix algebra} (Academic Press, New York, 1965).\\
 T.D. Lee and G.C. Wick, Nucl. Phys. B {\bf 9} (1969) 209.\\
 F.G. Scholtz, H. B. Geyer and F.J.H. Hahne, Ann. Phys. {\bf 213} (1992) 74.
\item A. Mostafazadeh, J. Math. Phys. {\bf 43} (2002) 205;{\bf 43} (2002) 2814;
{\bf 43} (2002) 3944.
\item Z. Ahmed, Phys. Lett. A {\bf 290} (2001) 19; {\bf 294} (2002) 287.
\item Z. Ahmed and S.R. Jain, ``Pseudo-unitary symmetry and the Gaussian pseudo-unitary
ensemble of random matrices", arXiv quant-ph/0209165 (also submitted to Phys.
Rev. E).\\
Z. Ahmed and S.R. Jain, ``Gaussian ensembles of $2 \times 2$ pseudo-Hermitian
random matrices" to appear in J. Phys. A: Math. Gen.  (2003, The special issue
on Random Matrices).\\
Z. Ahmed, ``An ensemble of non-Hermitian Gaussian-random $2\times 2$ matrices
admitting the Wigner surmise'' to appear in Phys. Lett. A (PLA 12155) 2003.
\item C.M. Bender, P.N.Meisinger and Q. Wang, ``All Hermitian Hamiltonians
have parity'', arXive quant-ph/0211123, J. Phys. A : Math. Gen. {\bf 36} (2003) 1029.
\item C.M. Bender, D.C.Brody and H.F.Jones, ``Complex extension of quantum
mechanics'', arXiv quant-ph/2010076, Phys. Rev. Lett. {\bf 89} (2002) 270401.
\item J.D. Bjorken and S. D. Drell, {\em Relativistic quantum fields}
(McGraw-Hill, New York,1965) ch. 15, pp. 107-123.
\end{enumerate}
\end{document}